\documentclass{chjaa}                    % preprint, the final version

\usepackage{graphicx}                   %for PS/EPS graphics inclusion
\input{epsfig1.sty}                        %for PS/EPS graphics inclusion

\begin{document}

\title{Strange stars: Can their crust reach the neutron drip density?}

\author{Hai Fu and Yong-Feng Huang}

\institute{Department of Astronomy, Nanjing University, Nanjing
210093, China \email{hyf@nju.edu.cn} }

\date{ChJAA in press}

\abstract {Electrostatic potential of electrons near the surface
of static strange stars at zero temperature is studied within the
frame of the MIT bag model. We find that for QCD parameters
within rather wide ranges, and if the nuclear crust on the strange
star is at a density leading to neutron drip, the electrostatic
potential is insufficient to establish an outwardly directed
electric field, which is crucial for the survival of such crusts.
If a minimum gap width of 200 fm is called in for a more stringent
constraint, our calculations completely rule out the possibility
of such crusts' presence on strange stars. Therefore, our results
prefer against the existence of neutron-drip crust in nature.
}

\authorrunning{Hai Fu \& Yong-Feng Huang}
\titlerunning{Strange Star Crusts}

\maketitle

%%%%%%%%%%%%%%%%%%%%%%%%%%%%%%%%%%%%%%%%%%%%%%%%%%%%%%%%%%%%%%%%
%%%%%%%%%%%%%%%%%%%%%%%%%%%%%%%%%%%%%%%%%%%%%%%%%%%%%%%%%%%%%%%%

\section{Introduction}

There may exist a more stable state of hadrons than $^{56}$Fe,
called strange quark matter (SQM), which is a bulk quark phase
consisting of roughly equal numbers of $u$, $d$, and $s$ quarks
plus a smaller number of electrons to guarantee charge neutrality
(Witten 1984). This theoretical result is unlikely to be proved or
disproved through QCD calculations at least in the foreseeable
future. Final adjudication must come from experiments conducted
with accelerators {\it or} from astrophysical tests. One of the
important consequences of Witten's hypothesis is the prediction of
strange stars (Haensel, Zdunik, \& Schaeffer 1986; Alcock, Farhi,
\& Olinto 1986, hereafter AFO), i.e., stars made of SQM. The
presence of such stars should never be rare, in fact some authors
even inferred that all neutron stars might have been converted to
strange stars, since the whole Galaxy is likely to be contaminated
by stranglets (Glendenning 1990; Madsen \& Olesen 1991; Caldwell
\& Friedman 1991; Medina-Tanco \& Horvath 1996). For decades,
unremitted efforts have been made to observationally discriminate
strange stars from neutron stars, and some candidates for strange
stars have been accumulating (Cheng et al. 1998; Li et al. 1999;
Xu et al. 2001a, b). However, even to now it is still premature to
reach any firm conclusions.

A strange star can have a normal matter crust. The density at the base of the
crust is a crucial parameter, which has been discussed by many authors
(AFO; Huang \& Lu 1997a, b; Xu \& Qiao 1999; Yuan \& Zhang 1999;
Chen \& Zhang 2001; Ma et al. 2002). Recently, Ma et al. (2002) studied
the influence of some parameters, such as the strange quark mass ($m_s$),
the bag constant ($B$), and the strong coupling constant ($\alpha_c$),
on the crust density.
However, their study does not give out clearly the ranges of parameters
that enable a crust as dense as neutron drip density. Here we study the
problem in more detail. The parameter ranges that support a neutron-drip-density
crust will be calculated numerically. Our paper is arranged as follows.
Section 2 describes the existence of strange star crust. Section 3 gives
the detailed procedure of our calculations and the numerical results in
the $\alpha_c - m_s - B$ space. Finally, we summarize
the results in Section 4.

\section{The Existence of Crust}

It was pointed out by AFO that a strange star could be covered by
a normal material crust. The presence of electrons in SQM is vital
to the existence of such crusts. Because $s$ quark's mass is
larger than $u$ and $d$ quarks, it is slightly deficient in
equilibrium SQM. Electrons are thus called in to make the system
electrically charge neutral. As quarks are bounded through strong
interaction, they should have a very sharp surface with thickness
of the order of 1 fm. On the other hand, the electrons, bounded by
the Coulomb force, can extend several hundred fermis beyond the
quark surface. So, a strong electric field, $\sim 10^{17}$ V
cm$^{-1}$ and outwardly directed, will be established in a thin
layer of several hundred fermis thick above the strange matter
surface. This field can support a crust, composed of normal
nuclear matter, suspended out of contact with the SQM core.

The so-called neutron drip density, $\rho_{drip} \sim
4.3\times10^{11}$g cm$^{-3}$ sets an absolute upper limit for the
density of the crust, $\rho_{crust}$. The reason is that,
if $\rho_{crust}$ reaches $\rho_{drip}$, neutrons will begin to
drip out. Being electrically neutral, the neutrons will fall freely
into the core, and, by
hypothesis, be deconfined to be SQM. As a consequence,
$\rho_{crust}$ will keep going down, until it's below
$\rho_{drip}$. Conventionally, when the nuclear crust is taken
into account, the bottom density was assumed to be $\rho_{drip}$.
However, is neutron drip the only limit on the crust density?
Could the crust density actually go so far? Huang \& Lu (1997a, b)
said NO to both questions. By proposing that mechanical balance
should be held between electric and gravitational forces on the
whole crust, not only on a single nucleus as modelled by former
authors (AFO), they claimed that at a density still far lower
($\sim \rho_{drip}/5$) than the neutron drip density, the crust
would begin to break down.

AFO proposed a model to describe the gap between the SQM core and
the nuclear crust within the framework of Thomas-Fermi model. The
electric field should be described with the following
Poisson's equation:
\begin{equation}
\frac{d^{2} V}{d z^{2}}=\left\{
     \begin{array}{llr}
       {4 \alpha (V^{3}-V_{\rm q}^{3})/[3 \pi (\hbar c)^2]}, & z \leq 0,  \\ \\
       {4 \alpha V^{3}/[3 \pi (\hbar c)^2]}, & 0 < z \leq z_{\rm G}, \\ \\
       {4 \alpha (V^{3}-V_{\rm c}^{3})/[3 \pi (\hbar c)^2]}, &  z_{\rm G} < z,
     \label{eq:alcock}
     \end{array}
   \right.
\end{equation}
where $z$ is a space coordinate measuring the height above the
quark surface, $\alpha$ is the fine-structure constant, $V_{\rm
q}^{3}/3\pi^{2} \hbar^3$ is the quark charge density inside the quark
matter, $V$ is the electrostatic potential of electrons,
$V_{\rm c}$ is the electron
Fermi momentum deep in the crust, which represents the positive
charge density of the ions within the crust,  and $z_{\rm G}$ is
the gap width between the SQM surface and the base of the crust.
In fact, $V_{\rm q}$ and $V_{\rm c}$ are the boundary values of
the above equation: $V \rightarrow V_{\rm q}$ as $z \rightarrow -
\infty$, $V \rightarrow V_{\rm c}$ as $z \rightarrow + \infty$.
Meaningful solutions to the above equations occur only if $V_{\rm
c} < V_{\rm q}$. Actually in a core-crust system, $V_q$ turns out
to be the electron electrostatic potential at $R_m$.
Here $R_m$ represents the maximum radius below
which electrical charge neutrality is locally satisfied
in the SQM core. Certainly $R_m$ is smaller than
$R$, the radius of the core.

Since the electron chemical potential (we will prove that it
equals to $V$ later) at which neutron drip occurs is $\sim$26
MeV (Baym, Pethick, \& Sutherland 1971), to keep a crust at a
density of $\rho_{drip}$ suspended, the electrostatic potential of
electrons near the edge of the SQM core, $V_{q}$, must at least
be larger than 26 MeV. In this work, we calculated $V_{q}$ for
static SQM cores at zero temperature. Because of the uncertainties
inherent in the critical QCD-related parameters, the $s$ quark
mass, $m_{s}$, the strong interaction coupling constant,
$\alpha_{c}$, and the bag constant, $B$, our calculations were
actually carried out in a space expanded by those three
parameters. We found that for parameters within rather wide
ranges, to support a crust at a density leading to neutron drip is
not possible. Especially, for the conventional choice of the
parameters, i.e., $m_s$=200 MeV, $\alpha_c$=0.3, and $B^{1/4}$=145
MeV, our calculations indicate $V_{q}\simeq$20 MeV, well below 26
MeV.

It is interesting to mention that the properties of charm-quark
stars (viz. strange-quark stars with an additional charm-quark
population) have been studied by Kettner et al. (1995). But
unfortunately, they found charm-quark stars are unstable against
radial oscillations, i.e., no such stars can exist in nature. So
our work will only pivot around strange-quark stars.

\section{Electrostatic Potential of Electrons in SQM Cores}

\subsection{Governing Equations}

The property of SQM is generally described using the
phenomenological MIT bag model (Chodos et al. 1974), which
simplifies the dynamics of confinement by introducing an
approximation that the quarks are separated from the vacuum by a
phase boundary and the region in which quarks live is endowed with
a constant universal energy density $B$. Since the description of
SQM has been introduced elsewhere (Farhi \& Jaffe 1984; Kettner et
al 1995), we will go straightforward to the governing equations and
describe the procedure of our calculations. Our goal is to
determine the electrostatic potential of electrons at $R_m$, i.e.,
$V_{q}$.

We assume that charge neutrality is locally satisfied. This is
true popularly: Due to the rearrangement of electron charge inside
and outside of the surface of an SQM core, the redundant positive
charge of the quarks will be balanced locally by $e^-$ only up to
radial distances $r\leq R_m$. However, $R_m$ in fact is only
minutely smaller (several hundred fermis, see AFO; Kettner et al
1995) than $R$, the core's radius.

Basing upon the realization that a core-crust system's temperature
is generally much smaller than the typical chemical potentials of
the constituents ($u, d, s, e^-$), we further assume the core is at
zero temperature. And we include first-order $\alpha_c$ effects in
our calculations.

The thermodynamic potentials (per unit volume) as functions of the
chemical potentials of the constituents read (Farhi \& Jaffe
1984;AFO):
\begin{eqnarray} % For 1-column style
\Omega_f(\mu_f)&=&-\frac{{\mu_f}^4}{4 \pi^2 (\hbar c)^3}
(1-\frac{2\alpha_c}{\pi}), ~~~ f= u, d, \label{eq:omegau} \\
\Omega_s(\mu_s)&=&-\frac{1}{4 \pi^2 (\hbar c)^3}\{\mu_s
({\mu_s}^2-{m_s}^2 c^4)^{1/2} ({\mu_s}^2-\frac{5}{2}{m_s}^2 c^4)+
\frac{3}{2}{m_s}^4c^8 {\rm ln}(\frac{{\mu_s}+({\mu_s}^2-{m_s}^2c^4)^{1/2}}{{m_s} c^2}) \nonumber \\
&&-\frac{2\alpha_c}{\pi}[3({\mu_s}({\mu_s}^2-{m_s}^2c^4)^{1/2}-{m_s}^2c^4{\rm ln}(\frac{{\mu_s}+
({\mu_s}^2-{m_s}^2c^4)^{1/2}}{{m_s}
c^2 }))^2-2({\mu_s}^2-{m_s}^2c^4)^2 \nonumber \\
&&-3{m_s}^4c^8{\rm ln}^2(\frac{{m_s} c^2}{{\mu_s}})
+6ln(\frac{\rho_R}{{\mu_s}})({\mu_s} {m_s}^2
c^4({\mu_s}^2-{m_s}^2c^4)^{1/2} \nonumber \\
&&-{m_s}^4c^8{\rm ln}(\frac{{\mu_s}+({\mu_s}^2-{m_s}^2c^4)^{1/2}}{{m_s}
c^2}))]\}\,, \label{eq:omegas} \\
\Omega_e(\mu_e)&=&-\frac{{\mu_e}^4}{12 \pi^2 (\hbar c)^3} \,.
\label{eq:omegae}
\end{eqnarray}
The renormalization point for the SQM, $\rho_R$, which appears in
Eq. (\ref{eq:omegas}), is chosen to be 313 MeV in this work,
because of the reasons pointed out by AFO. The number densities
for every constituent can be expressed in terms of $\mu_{i}$
($i=u, d, s, e$) from (Farhi \& Jaffe 1984;AFO)
\begin{equation}
n_{i}(\mu_{i})=-\frac{\partial\Omega_{i}}{\partial\mu_{i}} \,, \label{eq:ni}
\end{equation}
and charge neutrality requires
\begin{equation}
\frac{2}{3}n_{u}-\frac{1}{3}n_{d}-\frac{1}{3}n_{s}-n_{e}=0\,. \label{eq:neutrality}
\end{equation}

In fact, Eq. (\ref{eq:neutrality}) could be rendered to an
equation of $\mu_{i}$ ($i=u, d, s, e^-$) if we substitute Eqs.
(\ref{eq:omegau})-(\ref{eq:ni}) into it. Later in this section, we
will demonstrate that all $\mu_{i}$ at a given energy density can
be determined if combined with the chemical equilibrium conditions
and the energy density equation, and we will point out that
$V_{q}$ is equal to $\mu_e$.

Chemical equilibrium between the three quark flavors and the
electrons is maintained by weak interactions (i.e., $\beta$-stable
SQM, Farhi \& Jaffe 1984; AFO)
\begin{eqnarray}
d &\rightarrow& u + e + \overline{\nu}_{e}, \\
u + e &\rightarrow& d + \nu_{e}, \\
s &\rightarrow& u + e + \overline{\nu}_{e}, \\
u + e &\rightarrow& s + \nu_{e}\,,
\end{eqnarray}
and
\begin{equation}
s + u \leftrightarrow d + u\,.
\end{equation}
The loss of neutrinos by the star implies that their chemical
potential is equal to zero (we ignore the effect brought by the
finite mass of neutrinos). Hence, at equilibrium the chemical
potentials should obey:
\begin{equation}
\mu_{d}=\mu_{s} \equiv \mu\,, \label{eq:mudmus}
\end{equation}
\begin{equation}
\mu_{u} + \mu_{e}=\mu\,. \label{eq:muumue}
\end{equation}
Combined with the condition of charge neutrality (Eq.
(\ref{eq:neutrality})), these equations leave us with only one
independent chemical potential, which we have denoted by $\mu$
(see Eq. (\ref{eq:mudmus})). Therefore, we need still one equation
to make the system close.

The total energy density $\rho$ is given by
\begin{equation}
\rho = \sum(\Omega_i(\mu_i)+\mu_i n_i(\mu_i)) + B, ~~~ i=u, d, s,
e^-. \label{eq:energydensity}
\end{equation}
Since all $\mu_i$ could be expressed in terms of $\mu$, solving
Eq. (\ref{eq:energydensity}) can determine $\mu$ if given the
value of $\rho$, and so forth, $\mu_i$.

If the Thomas-Fermi model is invoked to describe the electrons
associated with the quarks, we could easily arrive at the
conclusion that $V = \mu_e$, i.e., the electrostatic potential
is equal to the chemical potential for the electrons (see, e.g.,
AFO; Kettner et al 1995).
The derivation is as follows.
The number density of electrons is given by the local Fermi
momentum $P_e$,
\begin{equation}
n_e={P_e}^3/(3\pi^2\hbar^3).
\end{equation}
On the other hand,
\begin{equation}
n_e=-\partial\Omega_{e}/\partial\mu_{e}={\mu_e}^3/(3\pi^2\hbar^3
c^3).
\end{equation}
From the above two equations, we get $\mu_e = P_e c$.
Since the electrons are confined within a sphere of infinite
radius, their total energy $-V+P c$ should obey, $-V+Pc \leq
-V(\infty)=0$, i.e., $-V+P_ec=0$, and so forth,
\begin{equation}
V=P_ec=\mu_e.
\end{equation}
As $\mu_e$ at $R_m$ could be calculated following the procedure stated
above, $V_q$ is readily at hand.

\subsection{Numerical Results}

Since $\mu_e$ decreases with density (Kettner et al 1995), which
reflects the fact that less electrons are needed in denser SQM,
the electrostatic potential of electrons increases monotonically
from the center toward the surface of strange stars. Because when
$r < R_m$ the electrical charge neutrality is locally satisfied,
only the electron electrostatic potential at $R_m$, corresponding
to $V_q$ in Eq. (\ref{eq:alcock}) is responsible
for supporting the nuclear crust. Owing to the equation of state
$P=(\rho-4B)/3$ (Witten 1984)\footnote{Although this EOS is
derived in the limit $m_s\rightarrow 0, \alpha_c\rightarrow 0$,
for intermediate values of $m_s$ this equation is less than 4\%
different from the full expression (AFO).}, the energy density at
the surface of SQM cores is universally equal to $4B$, independent
of the star's mass, or in other words, of the central density.
Hence, although in our calculations we fixed the energy density at
$4B$, our results will stand for the complete equilibrium
sequences of compact SQM core-crust systems as determined by
Glendenning, Kettner, \& Weber (1995).

There exist large uncertainties in the three QCD-related
parameters: the $s$ quark mass, $m_{s}$, the strong interaction
coupling constant, $\alpha_{c}$, and the bag constant, $B$. Our
calculations are actually carried out in a space expanded by
those parameters. The exact values of them are unknown but are
probably constrained within: 50 MeV $\leq m_s \leq$ 340 MeV, 0
$\leq \alpha_c \leq$ 0.6, and 135 MeV $\leq B^{1/4} \leq$ 165 MeV.

The contours at fixed $V_{q}$ in the $\alpha_c - m_s$ plane for
$B^{1/4}$= 135, 145, 155, 165 MeV are respectively presented in
Fig. 1(a)-(d). The solid and dashed curves refer to $V_{q}$=26 MeV
(the electron chemical potential at which neutron drip occurs) and
$V_{q}$=0 MeV, respectively. The gray regions show where the
energy per baryon of the SQM, $\mu_n$
($\mu_n \equiv \mu_u+\mu_d+\mu_s$), exceeds the lowest energy per baryon
found in nuclei, which is 930 MeV for $^{56}$Fe. The black regions
are regions of no physical solutions.

To support a crust at a density leading to neutron drip, $V_{q}$
must, at least, be larger than 26 MeV (in order to establish an
outwardly directed field, or in other words, to get meaningful
solutions to Eq. (\ref{eq:alcock})). So, only parameters in
regions above the solid curves can make such crusts possible, and,
of course, they should avoid the gray and black regions.
Conventionally, the three parameters are chosen to be $m_s$=200
MeV, $\alpha_c$=0.3, and $B^{1/4}$=145 MeV. The cross in Fig.
1({\it b}) represents this set. Obviously, it lives outside the
permitted region, and actually $V_{q}$ equals roughly 20 MeV.
We point out emphatically that $V_{q} < 26$ MeV here.

Fig. 1 clearly shows that for QCD parameters within rather wide
ranges, owing to the insufficient electrostatic potential of
electrons, an SQM core is not capable of carrying a
crust at $\rho_{drip}$ suspended.

Another factor we should take into account is the gap width, $z_G$.
Note that the lattice spacing in the crust is $\sim$200 fm, in the
same order of $z_G$, while the model (i.e., Eq. (\ref{eq:alcock}))
assumed a smooth distribution of the ionic charges in the crust.
In order to make their analysis self-consistent, AFO assumed
that the width of the gap should be larger than 200 fm. $z_G$
is entirely determined by $V_c$ and $V_q$ in AFO's model, we thus
can also determine $V_q$ if given the values of $z_G$ and $V_c$.
If $z_G$= 200 fm and $V_c$= 26 MeV (viz. a neutron-drip crust),
then $V_q$ was found to be $\sim$125 MeV after solving Eq.
(\ref{eq:alcock}) with the associated boundary conditions (i.e.,
$V \rightarrow V_{\rm q}$ as $z \rightarrow - \infty$, $V
\rightarrow V_{\rm c}$ as $z \rightarrow + \infty$, and $V, dV/dz$
both continuous at z=0 and at z=$z_G$) as well as the
approximation $dV/dz=0$ for $z\geq z_G$ (viz. the gravitation
forces are neglected\footnote{Though this approximation has been
criticized by Huang \& Lu (1997), it is accurate enough for our
purpose here.}). Therefore, if a gap width larger than 200 fm is really a
necessary qualification for the stability of a core-crust system,
the assumed presence of crusts at neutron drip density would have
been completely ruled out, since no QCD parameters can make
$V_{q}$ so high, at least within the ranges in which our
calculations were done.

Although our calculations were limited to SQM at zero temperature,
we can state unambiguously that a finite temperature can only make
such dense crusts' situation even worse. According to Kettner et
al. (1995)'s calculations (see Fig. 12 in their paper), $V_{q}$
decreases with temperature of the SQM. It thus becomes even more
unlikely that a neutron-drip crust can exist above strange stars
at higher temperatures.

\section{Conclusions}

We have studied the electrostatic potential of electrons near the
surface of SQM cores in a parameter space expanded by $m_s,
\alpha_c$ and $B$, in order to evaluate the feasibility of the
existence of crusts at neutron drip density on strange stars. Our
numerical results indicate:

1. For QCD parameters within rather large ranges (Fig. 1), the
electrons in an SQM core are incapable of establishing an
outwardly directed electric field to carry a neutron-drip crust
suspended.

2. If it is a sound criteria that the gap width $z_G$ should be
larger than 200 fm, then the possibility of neutron-drip crust's
presence on strange stars is completed ruled out by our
calculations.

Therefore, our results prefer against the presence of such crusts
in nature.

\vspace{1cm}

\noindent
{\bf Acknnowledgements} This work was supported by The Foundation
for the Author of National
Excellent Doctoral Dissertation of P. R. China (Project No: 200125),
the Special Funds for Major State Basic Research Projects, and the National
Natural Science Foundation of China.

%\newpage
\begin{figure}
\epsfig{file=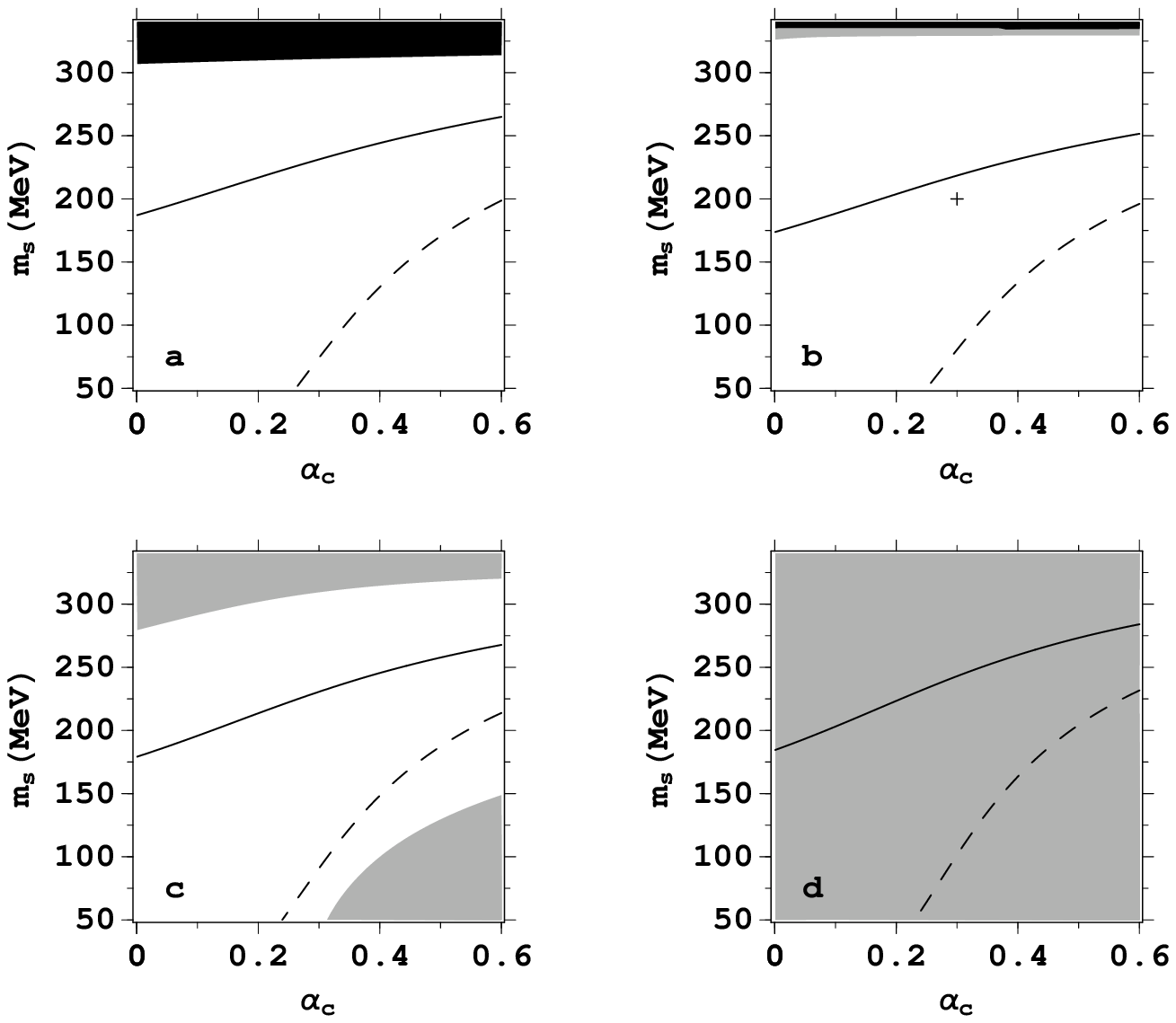, angle=0, height=110mm, width=11.5cm,
bbllx=80pt, bblly=50pt, bburx=440pt, bbury=410pt} \\
\caption{Contours of fixed
$V_{q}$ in $\alpha_c-m_s$ plane for $B^{1/4}=135, 145, 155, 165$
MeV. The cross near the center of {\it(b)} marks the position of
conventional choice of QCD parameters, i.e., $m_s$=200 MeV,
$\alpha_c$=0.3, and $B^{1/4}$=145 MeV. The solid and dashed curves
refer to $V_q$=26 MeV (the electron chemical potential at which
neutron drip occurs) and $V_q$=0 MeV, respectively. The gray
regions show where the energy per baryon of the SQM exceeds the
lowest energy per baryon found in nuclei, which is 930 MeV for
iron. The black regions are regions of no physical solutions.
Obviously, for QCD parameters within rather wide ranges (including
the conventional choice), SQM cores are incapable of supporting
crusts at neutron drip density (see text for details). These
results are independent of stellar mass, or in other words, of
central density, because in the MIT bag model the energy density
at zero pressure (corresponding to the surface of SQM cores) is
universally equal to $4B$. } \label{fig1}
\end{figure}


\begin{thebibliography}{}
\bibitem{} Alcock C., Farhi E., Olinto A., 1986, ApJ, 310, 261 (AFO)
\bibitem{} Baym G., Pethick C. J., Sutherland P. G., 1971, ApJ, 170, 299
\bibitem{} Caldwell R. R., Friedman J. C., 1991, Phys. Lett. B, 264, 143
\bibitem{} Chen C. X., Zhang J. L., 2001, Chin. Phys. Lett., 18, 145
\bibitem{} Cheng K. S., Dai Z. G., Wei D. M., Lu T., 1998, Science, 280, 407
\bibitem{} Chodos A., Jaffe R. L., Johnson K., Thorn C. B., Weisskopf V. F., 1974, Phys. Rev. D, 9, 3471
\bibitem{} Farhi E., Jaffe R. L., 1984, Phys. Rev. D, 30, 2379
\bibitem{} Glendenning N. K., 1990, Mod. Phys. Lett. A, 5, 2197
\bibitem{} Glendenning N. K., Kettner Ch., Weber F., 1995, ApJ, 450, 253
\bibitem{} Haensel P., Zdunik J. L., Schaeffer R., 1986, A\&A, 160, 121
\bibitem{} Huang Y. F.,  Lu T., 1997a, Chin. Phys. Lett., 14, 314
\bibitem{} Huang Y. F.,  Lu T., 1997b, A\&A, 325, 189
\bibitem{} Kettner Ch., Weber F., Weigel M.K., Glendenning N.K. 1995, Phys. Rev. D, 51, 1440
\bibitem{} Li X.-D., Bombaci I., Dey M., Dey J., vandenHeuvel E. P. J., 1999, Phys. Rev. Lett., 83, 3776
\bibitem{} Ma Z. X., Dai Z. G.,  Huang Y. F., Lu T., 2002, ApSS, 282, 537
\bibitem{} Madsen J., Olesen M. L., 1991, Phys. Rev. D, 43, 1069; erratum: 44, 4150
\bibitem{} Medina-Tanco G. A., Horvath J. E., 1996, ApJ, 464, 354
\bibitem{} Witten E., 1984, Phys. Rev. D, 30, 272
\bibitem{} Xu R. X., Qiao G. J., 1999, Chin. Phys. Lett., 16, 778
\bibitem{} Xu R. X., Xu X. B., Wu X. J., 2001a, Chin. Phys. Lett., 18, 837
\bibitem{} Xu R. X., Zhang B., Qiao G. J., 2001b, Astropart. Phys., 15, 101
\bibitem{} Yuan Y. F., Zhang J. L., 1999, A\&A, 344, 371

\end{thebibliography}
\end{document}